\documentclass{Interspeech2024}

\interspeechcameraready
\usepackage{amsmath,graphicx}
\usepackage{subcaption}
\usepackage{amsfonts} 
\usepackage{xcolor}
\usepackage{enumitem}
\usepackage{algorithm2e}
\usepackage{enumitem}

\usepackage{graphicx}
\usepackage{subcaption}

\setlist{nolistsep}
\usepackage{booktabs}
\usepackage{multirow}
\usepackage{hyperref}

\title{SeMaScore : A new evaluation metric for automatic speech recognition tasks}

\name{Zitha}{Sasindran}
\name{Harsha}{Yelchuri}
\name{T. V.}{Prabhakar}

\address{
  Department of Electronic Systems Engineering \\
  Indian Institute of Science, Bengaluru, India}
\email{\{zithas, harshay, tvprabs\}@iisc.ac.in}


\begin{document}

\maketitle

\begin{abstract}

In this study, we present SeMaScore, generated using a segment-wise mapping and scoring algorithm that serves as an evaluation metric for automatic speech recognition tasks. SeMaScore leverages both the error rate and a more robust similarity score.  We show that our algorithm's score generation improves upon the state-of-the-art BERTScore. Our experimental results show that SeMaScore corresponds well with expert human assessments, signal-to-noise ratio levels, and other natural language metrics. We outperform BERTScore by 41x in metric computation speed. Overall, we demonstrate that SeMaScore serves as a more dependable evaluation metric, particularly in real-world situations involving atypical speech patterns.

\end{abstract}

\section{Introduction and motivation}
\label{sec:intro}


The evaluation of automatic speech recognition (ASR) systems has predominantly centered around the utilization of error-based metrics, such as the word error rate (WER) and character error rate (CER). 
WER, in particular, has evolved as the de facto metric for benchmarking the performance of ASR models. While state-of-the-art (SOTA) ASR models trained on benchmark speech datasets have consistently demonstrated lower WERs, a notable challenge arises when these well-performing models assess noisy or disordered speech corpora \cite{disordered}, resulting in elevated WER values. Some examples include  ASR systems used in real-time applications like  medical transcription services applied to patients with speech-impairment or virtual assistants located in highly noisy environments. In these cases the focus needs to shift from quantifying errors to a more thorough assessment of whether the generated hypotheses accurately convey the intended meaning of the spoken sentence.

While WER serves as a valuable metric for quantifying word-level errors in ASR systems, it has limitations \cite{wernotgood,wer_alternative1} in capturing crucial aspects of ASR evaluation , such as semantic similarity and the significance given to each word error in the hypothesis. Consider two scenarios where the WER remains the same: one involving hyphenated words (well-being) represented as a single word (wellbeing) and the other involving misspelt word (velbing). Despite WER being the same in both cases, the former is more closer to the ground truth semantically. This observation implies that generating a hypothesis that perfectly aligns with the ground truth may be of little importance when comprehending the user's intent and the underlying semantic content. 


\begin{table}[]
\caption{Examples of cases where BERTScore fails: ground truth (GT) and hypothesis (H) generated.}
\label{tab:bertscore_failing}
\centering
\setlength{\tabcolsep}{2pt}
{\renewcommand{\arraystretch}{1.2}
\resizebox{0.46\textwidth}{!}{%

\begin{tabular}{cccc}
\hline
\textbf{GT} &\multicolumn{1}{c}{\textbf{H}} & \multicolumn{1}{c}{\textbf{BERTScore}} & \multicolumn{1}{c}{\textbf{SeMaScore}} \\ \hline
Smoking & Something        & 0.98 &0.3\\
Thank you lord & Thank you thank &\\
&thank thank  lord       & 0.89 &0.62\\
\hline

\end{tabular}

}
}
\end{table}

\begin{table*}[]
\caption{Example of segment mapping between the ground truth (GT) and hypothesis (H) generated.}
\label{tab:mapping}
\centering
{\renewcommand{\arraystretch}{1.2}
\begin{tabular}{ll}
\hline
\multicolumn{1}{c}{\textbf{Before mapping}} & \multicolumn{1}{c}{\textbf{After mapping}} \\ \hline
GT: I want to have a sandwich          & I | want | to | have a | sandwich \\
H: I vant to havea sand wich          & I | vant | to | havea | sand wich\\
\hline
\end{tabular}
}
\end{table*}

Over the years, numerous research works have sought to provide alternatives to WER \cite{inf_retrieval,weighted_wer2} as an evaluation metric and have demonstrated its limited alignment with human judgment when applied in the context of spoken conversations \cite{weighted_wer1}. Recent studies \cite{semantic_distance1, semantic_distance2,evaluating,heval} have introduced the utilization of pre-trained BERT-based \cite{bert} models to compute a semantic distance metric between the ground truth and hypothesis. However, the semantic distance metric calculated at sentence-level overly prioritizes some words over others, making it unreliable. In BERTScore \cite{bertscore}, the authors calculated the semantic distance at token-level as an evaluation criterion for ASR systems. However, BERTScore has some limitations as well, and this is presented in Table \ref{tab:bertscore_failing}. Table \ref{tab:bertscore_failing} presents two cases where BERTScore fails to provide an accurate score. The first example illustrates two entirely mismatched words, while the second example showcases a sentence with repetitions—a common mistake in ASR due to encoder-decoder model hallucination issues. In both instances, BERTScore assigns a higher similarity score and fails to detect the errors, despite the sentences being flawed. 
The authors identified this nature of BERTScore and proposed baseline rescaling, which tones down the value of BERTScore. However, we observed that some of the examples still fail after rescaling. For instance, if we consider the same example 1 of Table \ref{tab:bertscore_failing}, even after rescaling, the BERTScore drops marginally from 0.98 to 0.86. Further, baseline rescaling involves the computation of BERTScore over a corpus of million sentences for every sub-task in every language.
Motivated by these insights, we created a new metric that focuses on both the error rate and a more robust semantic similarity score for the comprehensive evaluation of ASR models.


In this paper, we introduce \textbf{SeMaScore}, a evaluation metric for ASR tasks. This innovative metric integrates traditional error rate-based metric with a robust semantic similarity-based approach to evaluate ASR outputs. 
Furthermore, our findings illustrate that SeMaScore performs effectively even when WER is considerably high when ASR models encounter atypical speech patterns such as disordered, noisy, or accented speech utterances. Our work demonstrates the robustness of the proposed metric in handling transcription errors and its strong alignment with expert human assessments, and other natural language metrics.  We show that other metrics fail to correlate well with some scenarios, whereas SeMaScore aligns well with almost all the scenarios. This highlights the versatility and reliability of our metric across a broad spectrum of ASR applications.


\section{SeMaScore}
\label{sec:sema_score}
We discuss the methodology for calculating the SeMaScore metric.
Our approach consists of four primary phases: Segment mapping, computing segment scores \& importance weights, and finally, the calculation of the SeMaScore.
\subsection{Segment mapping}
\label{sec:segment_mapping}


Utilizing contextual embeddings of complete sentences from the BERT model to compute sentence similarity, as outlined in \cite{semantic_distance1,semantic_distance2}, naturally prioritizes the keywords within each sentence. BERTScore improves upon this and takes a more complex approach by calculating cosine similarity between every token in both the ground truth (GT) and hypothesis (H), thus finding the mapping between the GT and H. While this token-level granularity is suitable for tasks like summarization and text generation, its application in ASR evaluation is computationally expensive.

ASR evaluation, being a sequential task, does not require the complex token-level mapping used in BERTScore. In our approach, we adopt a sentence alignment technique based on the Levenshtein distance to create a mapping between GT and H.
To find the mapping between the sentences, we first align GT and H based on the character-level edit distance. Subsequently, we identify corresponding words or groups of words (segments) between GT and H that align with each other. Table \ref{tab:mapping} provides an example of this segment mapping approach, where aligned portions of sentences correspond to single words or groups of words.
As demonstrated in the table, our approach accommodates cases where words are split (e.g., `sandwich' $\rightarrow$ `sand wich'), combined (e.g., `have a' $\rightarrow$ `havea'), or misrecognized (e.g., `have' $\rightarrow$ `hav'). This adaptable mapping strategy allows us to precisely identify the segment-wise correspondence ($GT_M$ and $H_M$), between GT and H.

\subsection{Computing segment score}
\label{sec:segment_score_calculation}
After obtaining the mapped segments, $GT_M$ and $H_M$, the next step is to assess their similarity. In our approach, we extract the token embeddings for each segment in $GT_M = \{GT_M[1],GT_M[2] \cdots GT_M[L]\}$ and $H_M =\{H_M[1],H_M[2]$ $\cdots H_M[L]\}$, from the contextual embeddings of GT and H, i.e., $e_{GT}$ and $e_H$.  Further, we use the mean pooling technique on the token embeddings of a segment to compute a single contextual embedding for each segment $i$ in $GT_M[i]$ and $H_M[i]$ to obtain  $e_{GT_M}[i]$ and $e_{H_M}[i]$.  Thereby, we ensure that the context of the segments is also considered.     

As explained in Section \ref{sec:intro}, by limiting to cosine similarity between token-level embeddings, as done in BERTScore, has its disadvantages of assigning a higher score. Therefore, we propose the introduction of a penalty for the cosine similarity score by multiplying it with (1 - match error rate (MER)). We leverage MER \cite{weighted_wer1}, an error-based metric used to assess character-level mistakes. MER is used instead of the traditional CER, as MER is bounded between 0 and 1.



\RestyleAlgo{ruled}

\SetKwComment{Comment}{/* }{ */}

\begin{algorithm}[t!]
\linespread{1.2}\selectfont
\caption{SeMaScore}
\label{alg:SeMaScore}
\KwData{Ground truth ($GT$), Hypothesis ($H$)}
\KwResult{$SeMaScore$}
    $GT_{M}, H_{M} \gets ~$Get\_Mapped\_Sentences$(GT, H)$\;
    $e_{GT}, e_{GT_{M}}, e_{H_{M}} \gets~$ Get\_Segment\_Embeddings$(GT, H, GT_{M}, H_{M})$\;    
    \For{$i \gets 1$ \KwTo len$(GT_M)$}{
    $SS_i \gets ~$ Cos\_Sim$(e_{GT_{M}[i]},e_{H_{M}[i]})$\;
    $SegScore_i \gets SS_i \times (1 - $MER$(GT_{M}[i],H_{M}[i]))$\;
    $\alpha_i \gets ~$Cos\_Sim$(e_{GT_{M}[i]},e_{GT})$\;
    $SeMaScore \gets SeMaScore + \alpha_i\times SegScore_i$\;
    $\alpha_{sum} \gets \alpha_{sum} + \alpha_i$\;
    }
$SeMaScore \gets SeMaScore / \alpha_{sum}$\;
\end{algorithm}


\subsection{Computing importance weights}
\label{sec:compute_importance}
When evaluating a sentence, it is crucial to assign importance to each segment based on its contribution to the meaning of the sentence. For example, as evident in Table \ref{tab:mapping}, the word `sandwich' carries more significance than the word `a'. 
BERTScore addresses this issue by leveraging term frequency-inverse document frequency (tf-idf) to assign weights to words across the ground truth corpus. However, this approach has its limitations. Consider the word `can,' which can have multiple meanings depending on the context. In the sentence `I can open the can of soda,' `can' illustrates the importance of context in understanding its intended sense. Using tf-idf weights alone fails to capture this contextual information, potentially resulting in an inadequate evaluation of the sentence.
To effectively determine the importance $\alpha_i$ of segment $i$, we calculate the cosine similarity between the embedding of the segment ($e_{GT_M}[i]$) and the embedding of the ground truth ($e_{GT}$). This method allows us to consider the contextual relevance of each segment, providing a more accurate assessment of its contribution to the overall sentence meaning.



\subsection{Metric calculation}

Algorithm \ref{alg:SeMaScore} outlines the complete process for calculating SeMaScore. Initially, we apply the segment mapping technique outlined in Section 2.1 to derive the mapped ground truth ($GT_M$) and hypothesis ($H_M$). Then, we extract contextual embeddings for each segment and compute the segment score, which is a combination of the similarity score and (1-MER), as elaborated in Section 2.2.
Following this, we determine the importance of each segment of $GT_M$ as explained in Section \ref{sec:compute_importance}. Finally, the SeMaScore is computed utilizing all these components, as shown in the Algorithm \ref{alg:SeMaScore}.






\section{ASR Analysis and results}
\label{sec:ASR_analysis}
We discuss the diverse settings employed to evaluate our SeMaScore metric. We outline the multiple datasets utilized in our analysis and describe the comparison methodology and results. All results were evaluated against the current SOTA BERTScore. All contextual embeddings were obtained using pre-trained roberta-base model. Revisiting the examples in  Table \ref{tab:bertscore_failing} where BERTScore fails, we can see that SeMaScore evaluated the hypotheses better by penalizing them appropriately. 
All the codes, datasets employed, and results are available from the following link: \url{https://github.com/zenlab-edgeASR/SeMaScore}.

\begin{figure}[!h]
\centering
  \begin{subfigure}[b]{0.23\textwidth}
		\centering
		\includegraphics[width=\textwidth,height = 3.3cm]{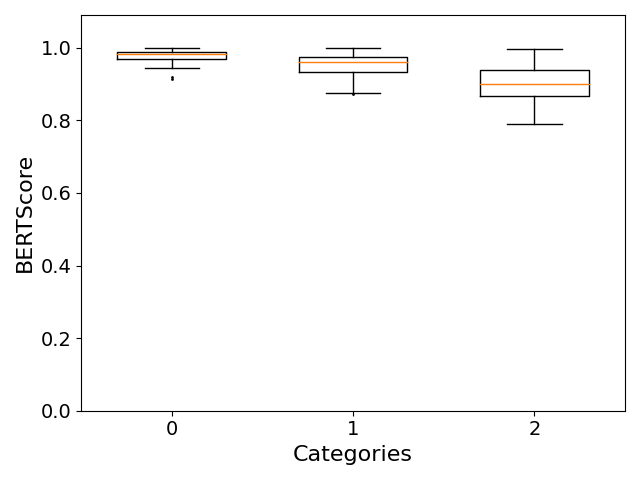}
         \caption{BERTScore}
         \label{fig:BERTSCore_torgo}
	\end{subfigure}
  \begin{subfigure}[b]{0.23\textwidth}
		\centering
		\includegraphics[width=\textwidth,height = 3.3cm]{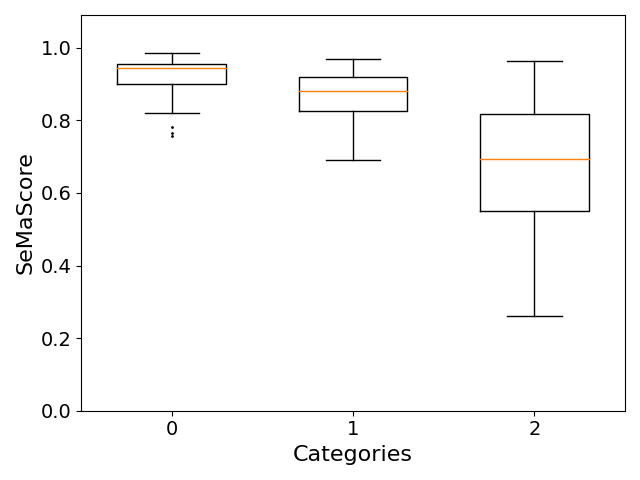}
         \caption{SeMaScore}
         \label{fig:SeMaScore_torgo}
	\end{subfigure}
        \caption{Distribution of various metrics for disordered speech}
        \label{fig:torgo}
\end{figure}


\begin{figure}[!h]
     \begin{subfigure}[b]{0.153\textwidth}
         \includegraphics[width=\textwidth,height = 3cm]{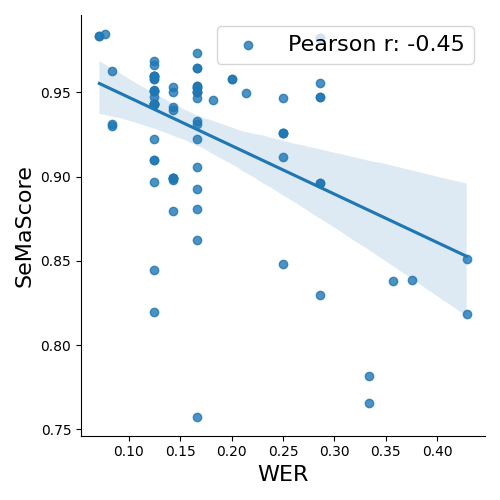}
         \caption{Category 0}
         \label{fig:corr_with_wer0}
     \end{subfigure}
     \begin{subfigure}[b]{0.153\textwidth}
         \includegraphics[width=\textwidth,height = 3cm]{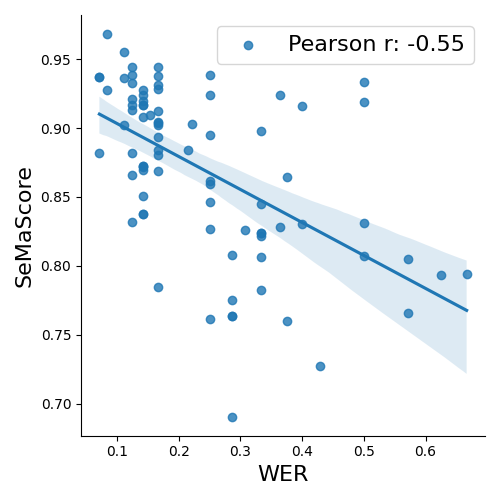}
         \caption{Category 1}
         \label{fig:corr_with_wer1}
     \end{subfigure}
     \begin{subfigure}[b]{0.153\textwidth}
         \includegraphics[width=\textwidth,height = 3cm]{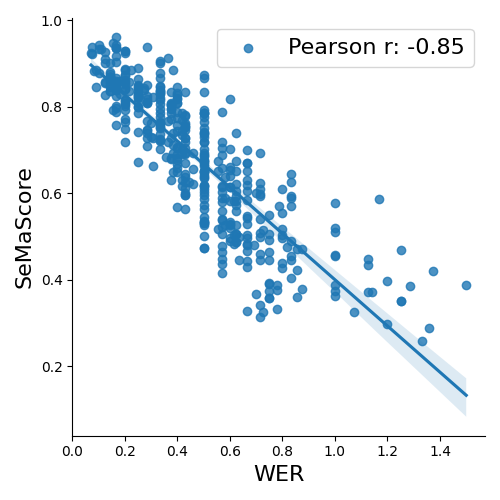}
         \caption{Category 2}
         \label{fig:corr_with_wer2}
     \end{subfigure}
        \caption{Correlation between SeMaScore and WER for disordered speech}
        \label{fig:disordered-correlation_with_wer}
\end{figure}

\begin{figure}[!h]
     \begin{subfigure}[b]{0.153\textwidth}
         \includegraphics[width=\textwidth,height = 3cm]{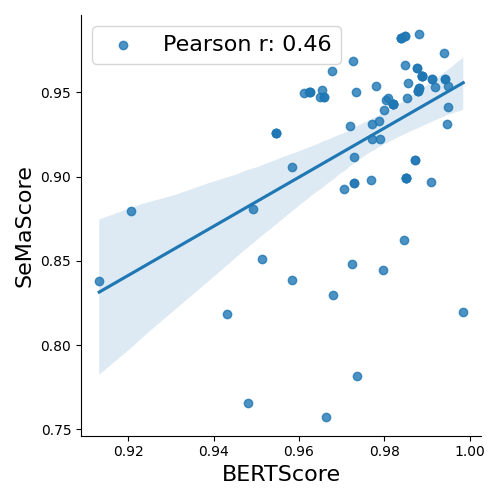}
         \caption{Category 0}
         \label{fig:corr_with_bertscore0}
     \end{subfigure}
     \begin{subfigure}[b]{0.153\textwidth}
         \includegraphics[width=\textwidth,height = 3cm]{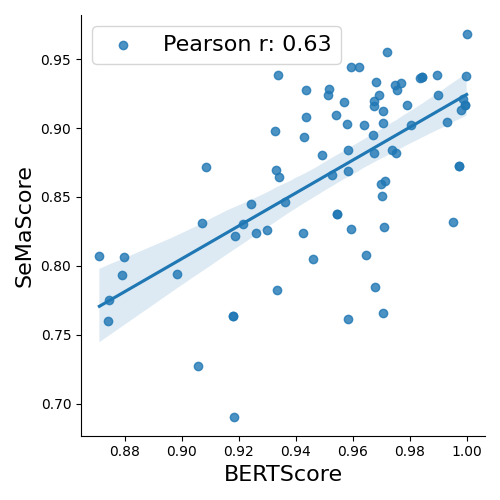}
         \caption{Category 1}
         \label{fig:corr_with_bertscore1}
     \end{subfigure}
     \begin{subfigure}[b]{0.153\textwidth}
         \includegraphics[width=\textwidth,height = 3cm]{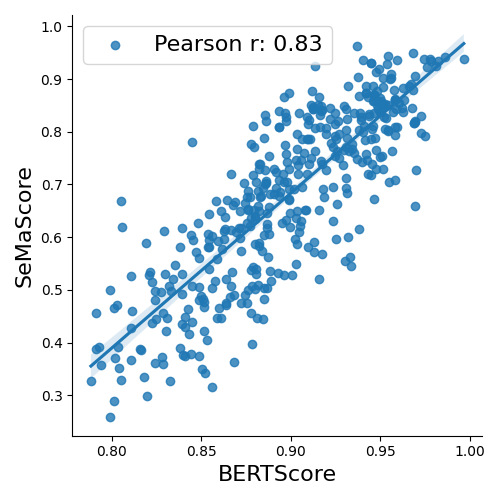}
         \caption{Category 2}
         \label{fig:corr_with_bertscore2}
     \end{subfigure}
        \caption{Correlation between SeMaScore and BERTScore for disordered speech}
    \label{fig:disordered-correlation_with_bertscore}
\end{figure}

\subsection{Disordered speech}
\label{sec:disordered_speech}
Our initial set of analyses aims to evaluate the effectiveness of our metric when applied to disordered speech datasets. To achieve this, we employ the Torgo dataset \cite{torgo}, which features speech recordings from individuals with dysarthria. We utilize two distinct models, namely DeepSpeech2 \cite{ds2} and Wave2vec2 \cite{wav2vec}, to generate inferences for 600 speech utterances (300 from each model) with an average WER of 0.4. In order to comprehensively assess the metric's performance and utility, we sought the expertise of two language specialists to measure the degree of similarity between the inferred sentences and the ground truth.
The assessment criteria we employ, as well as the observed error types, align with those described in \cite{bertscore_asr}. These assessments are categorized into three groups:
Category 0: meaning is entirely preserved, Category 1: meaning is mostly preserved, but some errors are present, and Category 2: meaning is entirely different.
We utilize this dataset to examine the correlation between the SeMaScore and human assessments.

Figure \ref{fig:torgo} presents the results obtained when BERTScore and SeMaScore are used to evaluate the disordered speech. It is clear from the figures that both BERTScore and SeMaScore correlate with human assessments.  From figures \ref{fig:BERTSCore_torgo} and \ref{fig:SeMaScore_torgo}, we can infer that for category 2, BERTScore has a mean of 0.9 with a standard deviation of 0.05 and SeMaScore has a mean of 0.68 with a standard deviation of 0.16 respectively. 
Since category 2 contains hypotheses whose meaning is different from that of the ground truths, we want the metrics to give a lower score which is well satisfied by SeMaScore.

Figure \ref{fig:disordered-correlation_with_wer} and Figure \ref{fig:disordered-correlation_with_bertscore}
depict scatter plots illustrating the correlation between SeMaScore and both WER and BERTScore. The plots show a negative correlation between SeMaScore and WER, as well as a positive correlation between SeMaScore and BERTScore across all categories. From Figure \ref{fig:disordered-correlation_with_wer}, it is apparent that the range of SeMaScore values corresponding to a specific WER value has a wide spread, indicating the consideration of semantic content. Figure \ref{fig:corr_with_bertscore2} demonstrates a strong correlation between our metric and BERTScore. However, SeMaScore produces a sensible range of scores (0.3 to 1) for categories containing semantically diverse sentences, compared to BERTScore (0.85 to 1).




\begin{figure}[!h]
  \begin{subfigure}[b]{0.23\textwidth}
		\centering
		\includegraphics[width=\textwidth,height = 3.3cm]{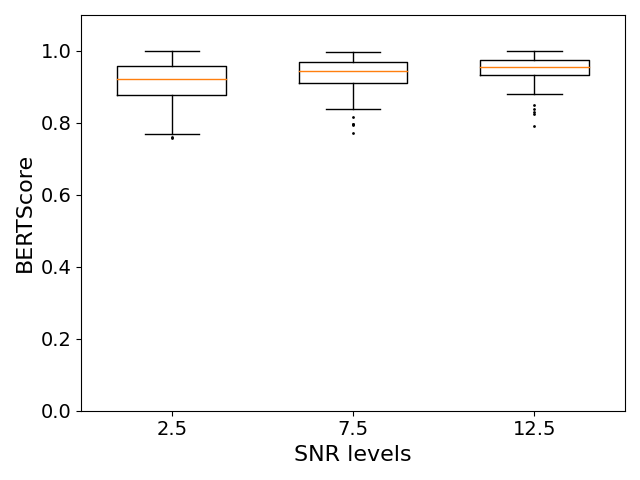}
         \caption{BERTScore}
         \label{fig:BERTSCore_voice_bank}
	\end{subfigure}
  \begin{subfigure}[b]{0.23\textwidth}
		\centering
		\includegraphics[width=\textwidth,height = 3.3cm]{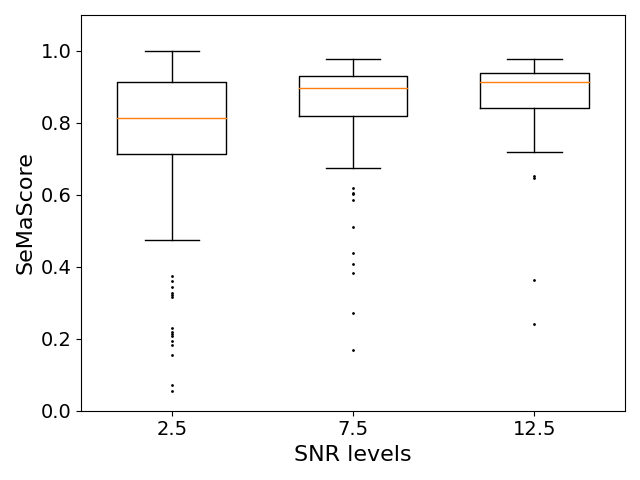}
         \caption{SeMaScore}
         \label{fig:SeMaScore_voice_bank}
	\end{subfigure}
        \caption{Distribution of various metrics for noisy speech}
        \label{fig:voicebank}
\end{figure}

\begin{figure}
    \centering
     \begin{subfigure}[b]{0.153\textwidth}
         \centering
         \includegraphics[width=\textwidth,height = 3cm]{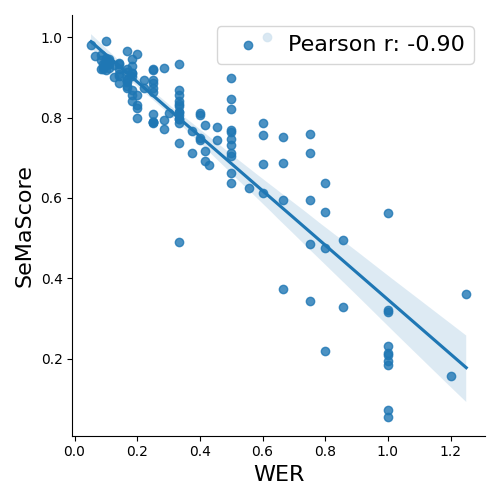}
         \caption{SNR level 2.5}
         \label{fig:corr_with_wer2.5}
     \end{subfigure}
     \begin{subfigure}[b]{0.153\textwidth}
         \centering
         \includegraphics[width=\textwidth,height = 3cm]{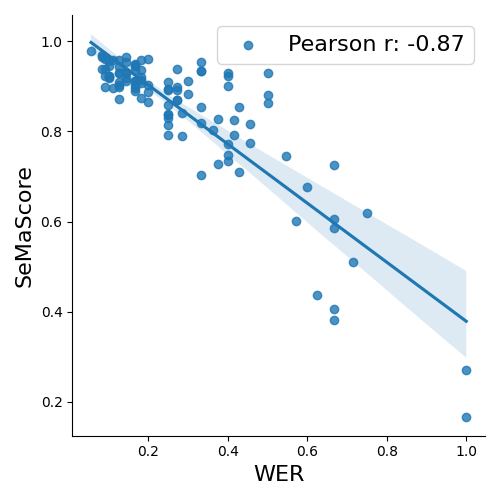}
         \caption{SNR level 7.5}
         \label{fig:corr_with_wer7.5}
     \end{subfigure}
     \begin{subfigure}[b]{0.153\textwidth}
         \centering
         \includegraphics[width=\textwidth,height = 3cm]{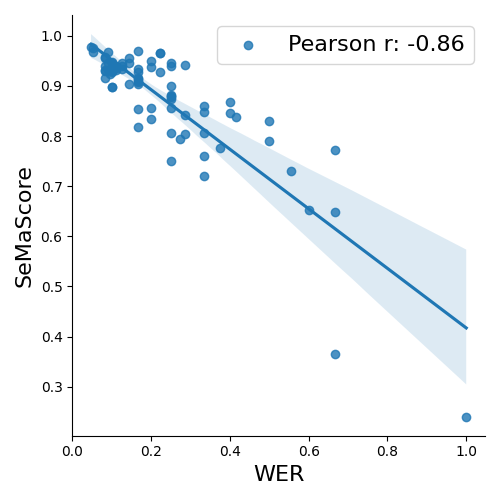}
         \caption{SNR level 12.5}
         \label{fig:corr_with_wer12.5}
     \end{subfigure}
        \caption{Correlation between SeMaScore and WER for noisy speech}
        \label{fig:noisy-correlation_with_wer}
\end{figure}

\begin{figure}
     \begin{subfigure}[b]{0.153\textwidth}
         \centering
         \includegraphics[width=\textwidth,height = 3cm]{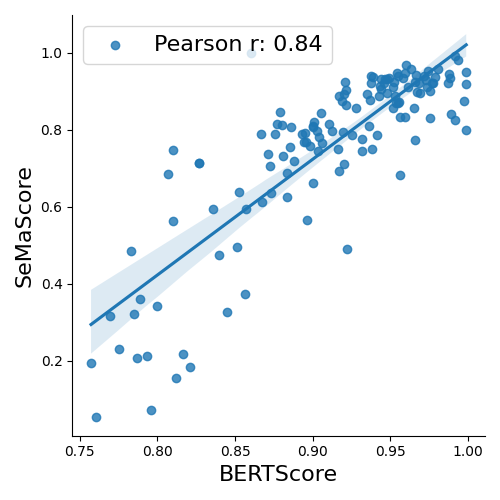}
         \caption{SNR level 2.5}
         \label{fig:corr_with_bertscore2.5}
     \end{subfigure}
     \begin{subfigure}[b]{0.153\textwidth}
         \centering
         \includegraphics[width=\textwidth,height = 3cm]{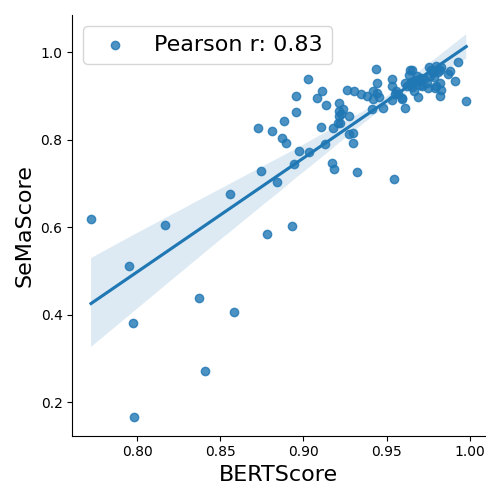}
         \caption{SNR level 7.5}
         \label{fig:corr_with_bertscore7.5}
     \end{subfigure}
     \begin{subfigure}[b]{0.153\textwidth}
         \centering
         \includegraphics[width=\textwidth,height = 3cm]{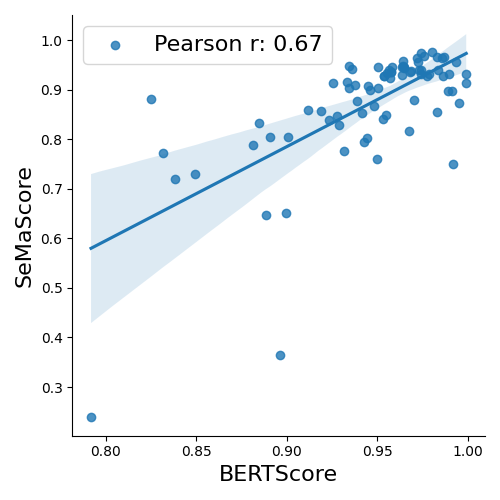}
         \caption{SNR level 12.5}
         \label{fig:corr_with_bertscore12.5}
     \end{subfigure}
        \caption{Correlation between SeMaScore and BERTScore for noisy speech}
        \label{fig:noisy-correlation_with_bertscore}
\end{figure}

\begin{table*}[!t]
\caption{Comparison of BERTScore and SeMaScore with intent accuracy and Named-Entity error rate}
\label{tab:nlp_tasks}
\centering
{\renewcommand{\arraystretch}{1.2}
\begin{tabular}{@{}cccccc@{}}
\toprule
\textbf{Datasets} & \textbf{WER}  & \textbf{BERTScore} & \textbf{SeMaScore} & \textbf{Intent Recognition Accuracy} & \textbf{Named-Entity Error} \\ 
                                   & \textbf{(\%)} &                                     &                                     & \textbf{(\%)}                        & \textbf{(\%)}          \\\midrule
Set A (ASR Outputs)                & 27                             & 0.94                               & 0.85                               & 88.16                                                 & 49.28                                        \\
Set B                              & 27                             & 0.94                               & 0.87                               & 90.62                                                 & 21.06                                        \\
Set C                              & 27                             & 0.90                               & 0.64                               & 81.25                                                 & 58.91                                        \\ \bottomrule
\end{tabular}
}
\end{table*}

\subsection{Noisy speech}
\label{sec:noisy_speech}
In this experiment, our objective is to investigate the effectiveness of the SeMaScore in evaluating speech utterances under noisy conditions. To achieve this, we make use of the voicebank-demand noisy speech testset \cite{voicebank}, which includes speech samples at various signal-to-noise ratio (SNR) levels, including 2.5dB, 7.5dB, and 12.5dB with an average WER of 0.31. We employ the Wave2vec2 model \cite{wav2vec} to generate inferences for these noisy speech utterances. Our primary goal is to assess how resilient our metric is to transcription errors in noisy speech and to examine its correlation with the SNR values.

Figure \ref{fig:voicebank} portrays the results discovered when BERTScore and SeMaScore are used to evaluate the noisy speech. The figures show that SeMaScore correlates well with SNR levels. We observe that, at a lower SNR level 2.5, the hypotheses generated correspond to assessment category 2. From figures \ref{fig:BERTSCore_voice_bank} and \ref{fig:SeMaScore_voice_bank}, we can infer that for SNR level 2.5, BERTScore has a mean of 0.9 with a standard deviation of 0.06 and SeMaScore has a mean of 0.71 with a standard deviation of 0.22 respectively. 
Since SNR level 2.5 includes hypotheses whose meaning differs from that of the ground truth, we expect the metrics to produce a lower score, which is again effectively delivered by SeMaScore.

Figure \ref{fig:noisy-correlation_with_wer} and Figure \ref{fig:noisy-correlation_with_bertscore} present scatter plots illustrating the correlation between SeMaScore with WER and BERTScore across various SNR levels. In Figure \ref{fig:corr_with_wer2.5}, at a lower SNR level of 2.5, where the hypotheses may contain more errors, SeMaScore values exhibit a broader distribution similar to WER. In Figure \ref{fig:corr_with_wer7.5} and Figure \ref{fig:corr_with_wer12.5} at higher SNR levels, both WER and SeMaScore values are confined to a lower range, as expected. Overall, WER and SeMaScore correlate well for this task.
Similar to our findings in disordered speech experiments, Figure \ref{fig:noisy-correlation_with_bertscore} showcases a strong correlation between our metric and BERTScore. SeMaScore produces scores within a broad range (0.1 to 1), in contrast to the higher BERTScore values (0.75 to 1).

\subsection{Domain-specific accented speech corpus}
\label{sec:atis_speech}
In this experiment, we conduct a comprehensive assessment of the SeMaScore in conjunction with other NLU-based metrics, specifically intent recognition accuracy and named-entity error rate, using a domain-specific accented speech dataset. To create this dataset, we curated 500 speech samples from the ATIS text corpus \cite{atis}. These speech samples were generated using a text-to-speech system \cite{speechelo} with five distinct accented voices. To evaluate our SeMaScore metric, we initially generated a hypotheses set (referred to as Set A)  by feeding these samples through the Deepspeech2 model. To further broaden our evaluation, we created two additional hypotheses sets, namely Set B and Set C, each containing the same WER as Set A.
Set B was constructed by introducing errors that have minimal impact on the sentence's overall meaning, such as inserting articles or splitting words in the reference ground truth. Conversely, Set C was generated by introducing errors that included removing or replacing keywords with random words. 


Table \ref{tab:nlp_tasks} illustrates the results obtained when we co-evaluate BERTScore and SeMaScore with other NLU-based metrics such as intent recognition accuracy and named-entity error rate. We can observe that BERTScore failed to correlate with both the intent recognition and named entity recognition tasks, whereas SeMaScore correlated with both of them. In case of intent recognition, SeMaScore is directly proportional to intent recognition accuracy, whereas it is indirectly proportional to named-entity error rate in the case of named entity recognition.





\subsection{Metric complexity}


Consider the scenario where GT consists of $m$ tokens and H contains $n$ tokens. If we consider the time required to calculate cosine similarity as $t_{cosine}$, then the computation time for the BERTScore metric can be expressed as $n\times m \times t_{cosine}$.
Let us assume the time required for edit-distance is given as $t_{edit}$. 
The metric calculation time for SeMaScore can be represented as $t_{edit}+ 2 \times max(n,m)\times t_{cosine}$ (worst case scenario).
We observed that the average metric calculation time for BERTScore on Torgo dataset took 1.95 seconds, whereas SeMaScore took 0.047 seconds. Thus, SeMaScore is 41 times faster than BERTScore.

\section{Summary and conclusions}
\label{sec:conclusions}

In this work, we introduce a novel evaluation metric for ASR tasks called  \textbf{SeMaScore},  which utilizes both error rate and a more robust similarity score to evaluate ASR outputs.
Our work compares SeMaScore with SOTA BERTScore and demonstrates that it aligns stronger than BERTScore in expert human assessments, SNR levels, and other natural language metrics. 
Moreover, our approach exhibits lower time complexity, resulting in metric computation times that is 41 times faster than that of BERTScore. These observations highlight the versatility and reliability of our metric across a wide range of ASR applications.

\newpage
\bibliographystyle{IEEEtran}
\bibliography{refs}

\end{document}